# Deep Learning in fNIRS: A review

**Condell Eastmond,[+*] Aseem Subedi*, Suvranu De, and Xavier Intes**
Department of Biomedical Engineering, Center for Modeling, Simulation and Imaging for Medicine (CeMSIM), Rensselaer Polytechnic, Department of Biomedical Engineering, 110 8th Street, Troy, NY, USA, 12180

**Abstract**.

**Significance:** Optical neuroimaging has become a well-established clinical and research tool to monitor cortical activations in the human brain. It is notable that outcomes of functional Near-InfraRed Spectroscopy (fNIRS) studies depend heavily on the data processing pipeline and classification model employed. Recently, Deep Learning (DL) methodologies have demonstrated fast and accurate performances in data processing and classification tasks across many biomedical fields.

**Aim:** We aim to review the emerging DL applications in fNIRS studies.

**Approach:** We first introduce some of the commonly used DL techniques. Then the review summarizes current DL work in some of the most active areas of this field, including brain-computer interface, neuro-impairment diagnosis, and neuroscience discovery.

**Results:** Of the 63 papers considered in this review, 32 report a comparative study of deep learning techniques to traditional machine learning techniques where 26 have been shown outperforming the latter in terms of classification accuracy. Additionally, 8 studies also utilize deep learning to reduce the amount of preprocessing typically done with fNIRS data or increase the amount of data via data augmentation.

**Conclusions:** The application of DL techniques to fNIRS studies has shown to mitigate many of the hurdles present in fNIRS studies such as lengthy data preprocessing or small sample sizes while achieving comparable or improved classification accuracy.

*These authors contributed equally to this work

[+]eastmc2@rpi.edu

**Introduction**

Over the last two decades, functional Near-InfraRed Spectroscopy (fNIRS) has become a well-established neuroimaging modality to monitor brain activity [1]. The ability of fNIRS to quantify cortical tissue hemodynamics over a long time, with relative high-spatial sampling and temporal resolution, has enabled its adoption in numerous clinical settings [2, 3]. fNIRS offers the unique advantage to be employed in freely mobile subjects with less restrictions than electroencephalography (EEG) or functional Magnetic

Resonance Imaging (fMRI). This permits the deployment of fNIRS in naturalistic scenarios and in patient populations that are typically not considered suitable for EEG or fMRI imaging [4]. Still, fNIRS faces numerous challenges for increased clinical adoption due to experimental settings [5], variations in statistical results [6], etc. Of importance, current trends in fNIRS aim to improve spatial resolution via increased spatial sampling, improve cortical sensitivity by using data processing to remove unwanted physiological noise, improve quantification by anatomical co-registration, and increase robustness via artifact identification and removal [7]. Current algorithmic implementations, however, require a high level of expertise to set up parameters that can be system and/or application-specific but also greatly impact the interpretability of the processed data. Moreover, the computational cost of these methods does not lend itself to bedside implementations. Following a ubiquitous trend in the field of data processing and analysis, new approaches leveraging developments in Deep Learning (DL) have been recently proposed to help overcome these caveats to a large extent.

The successes of DL methodologies across all biomedical engineering fields promise the development of dedicated data-driven, model-free data processing tools with robust performances, user-friendly employability and real-time capabilities. DL methods are increasingly utilized across the biomedical imaging field, including biomedical optics [8] and neuroimaging modalities including fMRI, magnetoencephalography (MEG) and EEG [9]. Following this trend, DL methodologies have also been recently used for fNIRS applications. In this review, we provide a summary of these current efforts. First, we introduce the basic concepts of DL including training and design considerations. Second, we provide a synthetic summary of the different studies reporting DL models in fNIRS applications. This section is divided into sub-field, namely Brain Computer Interface (BCI), clinical diagnostic and analysis of cortical activity. We then provide a short discussion and future outlook section.

**1. Deep Learning Methodology**

Deep learning can be viewed as black-box version of parametric machine learning techniques. Traditional machine learning techniques might make several assumptions about raw data distributions. Most notable is that data can be mapped to distinct classes (categorical data) or a regression line (continuous data) by a suitable transformation of input data. In parametric methods, weights are used to reduce multidimensional input data into separable space. These weights are learned by minimizing the objective function, trying to reduce error in prediction of score. With the introduction of hidden layers between input and output space, it is argued [10] that several abstractions can be learned that help in better distinction. These models are termed Artificial Neural Networks (ANN).

Inspired by biological neurons, ANNs generate a map between the input training data (x) and the output (y) using simple nonlinear operations performed at nodes, or "neurons", that form a computational graph [11]. The weights ($\Theta$) of the edges of the graph are updated, i.e., "trained" by minimizing a loss function $L(y, \hat{y})$ that measures the difference between the model output ($\hat{y}$) and the true output (y). Network training is accomplished efficiently using the chain rule of differentiation in an algorithm known as *backpropagation* using gradient descent [12]. The number of nodes in each layer of the graph defines the width of the network, whereas the number of layers defines its depth.

A deep network is one with sufficient depth, though there is no consensus on how deep the network has to be considered a "deep neural network". The entire network can be viewed as a differentiable function that learns a relationship between the input and the output using multiple levels of function composition, with the initial layers learning low-level features of the input, and the deeper layers extracting higher-level features. The advent of high-performance computing and the availability of large-scale data are fueling the current rapid advances in DL.

**1.1 Deep Learning Architectures**

Deep learning derives its versatility from the available cell/nodal operations. These operations include linear transformations, filters, and gates that have been inspired by other domain-specific tools, which compute abstract features in the hidden layers. In general, the choice of these operations defines the application of a network. The three most commonly used types of networks used in fNIRS studies are shown in **Figure 1**.

Dense networks like multi-layer perceptrons (MLPs) use linear transformation as cell operations and are synonymous to ANNs. Convolutional Neural Networks (CNNs) use convolution operations, where a fixed size filter is used for convolution over the input image or feature map. These are shown to be highly capable of recognizing digits [13], objects [14] and images in general. Similarly, Long-Short-Term-Memory (LSTM) networks can be unfurled in the time domain to learn time series data, including handwriting [15] and semantics for language translation [16]. LSTM cells use gates to maintain a cell-state memory [17]. Appropriate data are passed through the cells in successive timesteps, avoiding the vanishing gradient problem (see Sec 1.2.1) for long time series data [18].

A fundamental attribute of neural networks is introducing nonlinearities using so-called "activation functions" – based on the idea of the firing of biological neurons. The most common types of activation functions are listed in **Table 1**. The sigmoid function compresses values in the range of 0 and 1, activating large positive values while zeroing out large negative ones. The sigmoid function can also be used at the output for binary classifiers, for example in [19,20]. Dolmans [21] uses the sigmoid output for seven-class classification, although the softmax activation is widely used for multi-class classification. The softmax activation uses exponentiation followed by normalization to assign probabilities to the outputs. Higher softmax outputs at the output layer may be interpreted as confidence in the prediction [22].

The rectified linear unit (ReLU) activation is most widely used for the hidden nodes since it is computationally inexpensive and helps resolve the vanishing gradient problem [22]. It has proven to be efficient and effective for CNNs. It deactivates all nodes with negative outputs, which, although effective, deactivates those nodes throughout the training. This issue, also known as dead-ReLU problem, is solved by a recent activation function called exponential linear unit (ELU), which has also been used in Mirbagheri et al. [19], Saadati et al. [23], Ortega and Faisal [24], whereas some also use another variant called leaky-ReLU [25,26].

The final ingredient of neural networks is the loss function, which depends upon the output type. For classification tasks, the cross-entropy loss function [27] measures the difference in the probability distributions between ground truths and network predictions. Multi-class classifications [28-31] use categorical cross-entropy as the loss function, whereas binary classification problems use binary cross-entropy. For regression [32] or reconstruction [33] problems, the mean squared error loss function is used. These loss functions may be modified to implement constraints or regularization, e.g., a linear combination of MSE, variance, and two other metrics for a denoising autoencoder in [34].

**1.2 Practical Considerations in Deep Learning**
**1.2.1 Training Deep Networks**
After deciding on functions and architecture suitable to the problem, it is important to understand the underlying considerations involved to help the network learn the input-output map. It is common to normalize or scale the input to keep the parameters within tractable bounds. Two common methods include minmax scaling, where data is scaled between 0 and 1, and standardization, where data is scaled to have zero mean and unit variance. Though some networks have been shown to learn well without normalization [35] if the input data does not have a vast range, scaling is recommended to help convergence. Weight initialization also aids in convergence [36]. Techniques include the He and Glorot [36] initializations, where weights are drawn randomly from a normal or uniform distribution with predefined statistical moments. Moreover, dropout [37] is usually introduced between layers to randomly switch off a certain fraction of nodes so that the network does not overfit the training data.

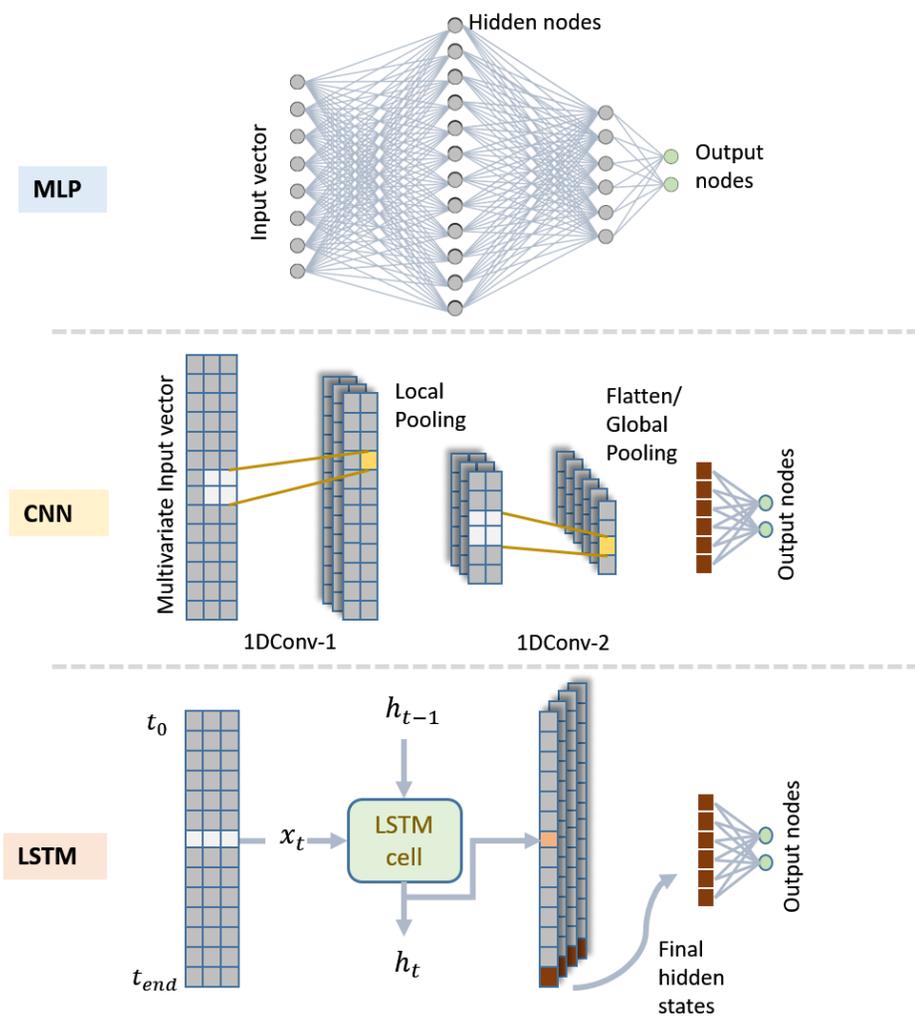

**Figure 1:** Illustrations of the three most common classes of network architectures used in the reviewed articles are shown in the figure above. (Top) A Multilayer Perceptron has all nodes fully connected. (Middle) A Convolutional NN (CNN) with a kernel size 2x2, with subsequent pooling layers. (Bottom) A Long Short Term Memory (LSTM) architecture where final hidden states are used. For illustrative purposes, the output layer is constructed for binary classification problems.

| Activation function | Output values | Bounds |
|---|---|---|
| **Sigmoid** | $\dfrac{1}{1+e^{-x}}$ | $(0,1)$ |
| **Softmax** | $\dfrac{e^{x_i}}{\sum_j e^{x_j}}$ | $(0,1)$ |
| **ReLU** | $max\,(0,x)$ | $[0,\infty)$ |
| **Leaky-ReLU** | $if\ x<0, \alpha x\ else\ x$ | $(-\infty,\infty)$ |
| **ELU** | $if\ x < 0, \alpha(e^x - 1)\ else\ x$ | $(-\alpha, \infty)$ |

**Table 1**: Activated outputs of input 'x' from each of the activation functions is shown in the table. Also shown are the bounds/range of activated values.

Gradient updates during *backpropagation* can either explode [38] or vanish if depth is too large [22] ReLU and LSTM cells were developed primarily to overcome this problem. Additionally, skip connections between deep layers help in propagating gradients backward, avoiding vanishing gradients [39]. These can be additive, e.g., ResNet [39], or augmentative, e.g., UNet [40]. On the other hand, gradient clipping and use of the SELU [38] activation function have been proposed to solve the exploding gradient problem.

Deep networks have multiple hyperparameters that are not actively learned or updated during training but set based on the literature or using a systematic search process. Weights are successively updated backward from the output layer by computing the mean gradients from a batch of samples. This batch size is a hyperparameter, commonly set to 32, although it depends on the architecture and hardware capabilities. Larger batch sizes demand more computation and memory for a single iteration, whereas smaller sizes imply slower convergence, r.g. Rojas et al. [41] use 64 batches in each update, whereas Wickramaratne et al. [42] use a batch size of 8. Another critical hyperparameter that controls the convergence rate is the learning rate that multiplies the loss-gradient in the gradient descent algorithm. Learning rate is usually set to values of the order of $10^{-2}$ at the start, and it can be decreased further for fine-tuning. For example, Ortega [24] uses an initial learning rate of 0.03, which decays at a factor of 0.9 after each epoch. An epoch is a point at which all non-overlapping batches in the dataset have been exhausted for training. The number of epochs, also another hyperparameter, is decided strictly based on when the network begins to converge. The most common optimizer algorithm for gradient descent is Adam [43] which uses an adaptive learning rate aided by momentum that helps network parameters to converge efficiently. Other algorithms also used in fNIRS applications are SGD [44] and RMSprop [42].

A notorious problem with training deep neural networks is the change in the distribution of inputs in every layer, which calls for careful initialization of weights, learning rate schedule, and dropout. Ioffe [45] introduced a technique called batch-normalization that helps mitigate this problem termed as 'internal covariance shift'. Batch normalization reduced training steps by a factor of 14 times in the original study while requiring less stringent conditions of initialization and learning rates. Hence, during batch training in CNN, it is often recommended to use batch normalization layers after each convolution [19,24].

Most biomedical applications have a limited number of subjects and limited training data (see Table 2 for a summary of the data set characteristics, including number of participants, number of channels and cortical areas monitored for all studies summarized herein). Hence, it is essential to take proper measures to avoid using a large network with a small dataset to prevent overfitting. The network will reduce the training error but lose the ability to generalize to unseen datasets. To mitigate this, it is good to start with the simplest network possible, have fewer weights, and iteratively improve the networks by adding the number of layers/nodes. Furthermore, overfitting can be avoided by regularizing weights (which adds a regularization loss to overall loss function), adding dropout layers, etc.

Since fNIRS data is sequential time series data, studies where segments of obtained data, such as resting state [46,47] are sufficient for analysis, utilize a sliding window approach. Here, a fixed-length data segment is extracted at fixed intervals, allowing overlap between subsequent windows. This is not applicable if the entire trial duration has to be analyzed, in which case each trial will have to be a single sample.

**1.2.2 Model Evaluation**

Although various modeling tools are available for the analysis of collected data in psychological and behavioral science, there is a growing concern about reproducibility of results using the settings reported in studies [48]. Although direct replication or even conceptual replication [49] might not be possible for many neuroimaging studies involving human subjects, studies now rely on simulated replication as the next best approach [50]. This entails partitioning the data into a number of subsets and assessing model performance on each of the subsets after it is trained with the rest of the data. Performance metrics on n-subsets a.k.a test sets, are averaged to get the mean performance metric, which is representative of the expected model performance on any unseen subset of data. Since the test set is "held out" from training, this method, also known as 'cross-validation' (CV) can be deemed as a crude measure of the generalizability of the model [50].

The type of subsets chosen depends on experimental settings, research question or the limitations of the dataset. The subsets selected can either be a shuffled subset of trials, one single trial, or, if available, trials of each participant subject. These CV schemes are called k-Fold CV, Leave-One-User-Out (LOUO) CV, and Leave-One-Subject-Out (LOSO) CV, respectively. Leave-One-Super-Trial-Out is another available rigorous CV technique [51]. Many of the reviewed papers carry out a 10-fold CV, whereas a few execute

Leave-One-Subject-Out CV [52,53]. The most common metrics used for evaluation are accuracy, specificity, and sensitivity.

### 1.2.3 Inputs to Deep Networks:

After the preprocessing steps (see Sec. 2.1), changes in oxy ($\Delta HbO$) and deoxy ($\Delta HbR$) hemoglobin, as well as total change ($\Delta HbT$) are obtained in time-series format. From these, data samples are segmented based on task settings. Data is segmented trialwise for task-based experiments, and for resting state data or long trials, data is segmented using sliding windows as mentioned previously.

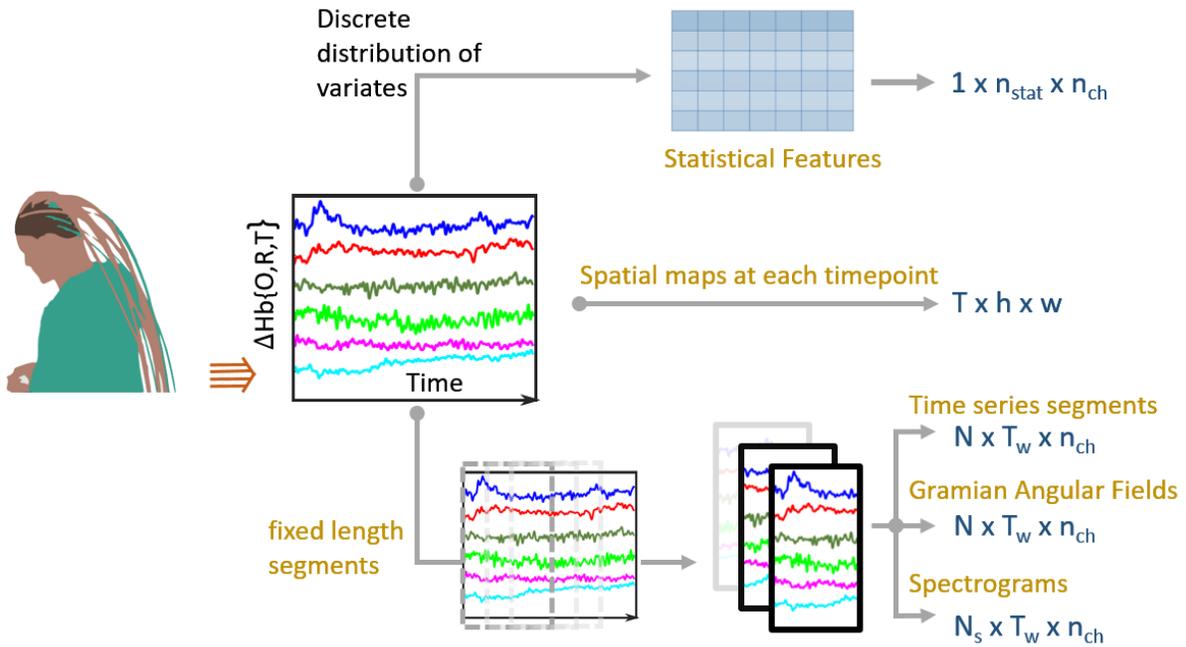

**Figure 2:** Extraction of samples to be used for training NNs. Time-series data denoting haemodynamic concentration changes are obtained from the raw data after initial analysis, and changed to appropriate formats (shown in right in the form of dimensions of input data samples), based on chosen architecture. N, Ns= Number of samples; T, Tw = Number of Timepoints; $n_{ch}$ = number of channels; h,w = height and width of spatial map images; $n_{stat}$= number of statistical moments/features

Many studies opt for the discrete probability distribution of concentration changes, and extract statistical features like mean, slope, variance, skewness, kurtosis, max and range in the form of manual features. In other cases, where the network is allowed to learn and extract features itself, the data is fed in forms of either spatial maps [20,28,54], or time series themselves. In some studies, segments of data are converted to other forms like Gramian Angular Fields [32] or Spectrogram Maps [30]. A schematic in **Figure 2** summarizes the sample extraction procedure employed in the studies. While the general trends and techniques used for deep learning fNIRS studies have been examined, the applications and some application-dependent techniques are further discussed in the next section.

| Citation No. | Author | Year | Number of Participants | Approximate Time Recorded per Participant | Number of Channels | Sampling Rate | Region of Brain | Dataset |
|---|---|---|---|---|---|---|---|---|
| | *Name of the Author* | *Published Year* | i.e. | i.e. 15-30 minutes | i.e. | Sampling Rate in Hz (Resampled Rate in Hz) | *Part of the brain being observed* | *Name of the public dataset* |
| 19 | Mirbagheri et al. | 2020 | 10 | 10 minutes | 23 | 10 | Prefrontal Cortex | N/A |
| 20 | Tanveer et al. | 2019 | 13 | 30 minutes | 28 | 1.81 | prefrontal & dorsolateral prefrontal cortex | N/A |
| 21 | Dolmans et al. | 2021 | 22 | 77-345 minutes | 27 | 10 | Not Specified | N/A |
| 23 | Saadati et al. | 2019 | 26 & 29 | 60-180 minutes | 24-36 | 10 | Not Specified | Simultaneous acquisition of EEG and NIRS during cognitive tasks for an open access dataset |
| 24 | Ortega & Faisal | 2021 | 12 | 13 Minutes | 24 | 12.6 | Sensorimotor Cortex | N/A |
| 25 | Dargazany et al. | 2019 | 10 | 12.5 minutes | 80 | 7.8125 | Motor Cortex | N/A |
| 26 | Nagasawa et al. | 2020 | 9 | 40 minutes | 41 | 10 | Sensorimotor Regions | N/A |
| 28 | Ghonchi et al. | 2020 | 29 | <30 minutes? | 36 | 10 (128) | Not Specified | Open Access Dataset for EEG+NIRS Single-Trial Classification |
| 29 | Trakoolwilaiwan et al. | 2017 | 8 | 1000 seconds | 34 | 25.7 | Motor Cortex | N/A |
| 30 | Janani et al. | 2020 | 10 | 60 minutes | 20 | 15.625 | Motor Cortex | N/A |
| 31 | Yoo et al. | 2021 | 18 | 19 minutes | 44 | 1.81 | Auditory Cortex | N/A |
| 32 | Gao et al. | 2020 | 13 | 3-10 hours | 12 | Not specified | Prefrontal Cortex | N/A |
| 33 | Ortega et al. | 2021 | 10 | 25 minutes | 24 | 12.5 | bilateral sensorimotor cortex | N/A |
| 34 | Gao et al. | 2020 | 30 | <25 minutes | 32 | Not specified | Prefrontal & Primary Motor Cortex & Supplementary Motor Area | N/A |
| 35 | Ma et al. | 2021 | 36 | 200 seconds | 31 | 7.14 | Not Specified | N/A |
| 41 | Fernandez Rojas et al. | 2021 | 18 | 3-5 minutes? | 24 | 10 | Somatosensory Cortex | N/A |
| 42 | Wickramaratne Mahmud | 2021 | 29 | <30 minutes? | 36 | 13.3 | Not Specified | Open Access Dataset for EEG+NIRS Single-Trial Classification |
| 44 | Sirpal et al. | 2019 | 40 | 30-180 minutes | Not Specified | 19.5 | bilateral anterior, middle & posterior temporal regions & frontopolar, frontocentral & dorsolateral frontal regions | N/A |
| 46 | Xu et al. | 2019 | 47 | 8 minutes | 44 | 14.29 | bilateral inferior frontal gyrus and temporal cortex | N/A |
| 47 | Yang et al. | 2020 | 24 | 27 minutes | 48 | 8.138 | Prefrontal Cortex | N/A |
| 52 | Takagi et al. | 2020 | 15 | 2 minutes | 22 | 10 | Prefrontal Cortex | N/A |
| 53 | Lu et al. | 2020 | 8 | 12-16 minutes | 52 | 10 | Prefrontal Cortex | Single-trial classification of antagonistic oxyhemoglobin responses during mental |
| 54 | Saadati et al. | 2019 | 26 | 33 minutes | 36 | 10.4 | Not Specified | Simultaneous acquisition of EEG and NIRS during cognitive tasks for an open access dataset |
| 56 | Benerradi et al. | 2019 | 11 | 30 minutes | 16 | 2 | Prefrontal Cortex | N/A |
| 60 | Liu et al. | 2021 | 18 | 6 minutes | 8 | 11.8 | Anterior Prefrontal Cortex | N/A |
| 61 | Lee et al. | 2018 | 6 | 30 minutes | 40 | Not specified | Supplementary Motor Area and Primary Motor Cortex | N/A |
| 62 | Kim et al. | 2022 | 42 | 540 seconds | 8 | 10 | Bilateral Prefrontal Areas | N/A |
| 63 | Wickramaratne & Mahmud | 2021 | 30 | 50 minutes | 20 | Not specified | Motor Regions | Open-Access fNIRS Dataset for Classification of Unilateral Finger- and Foot-Tapping |
| 64 | Woo et al. | 2020 | 11 | 7 minutes | 36 | 11 | Left Motor Cortex | N/A |
| 65 | Hennrich et al. | 2015 | 10 | 37 minutes | 8 | 10 | Prefrontal cortex | N/A |
| 66 | Kwon & Im | 2021 | 18 | 12 minutes | 16 | 13.3 | Prefrontal Cortex | N/A |
| 67 | Ho et al. | 2019 | 16 | 90 minutes | 7 | 18 | Prefrontal Cortex | N/A |
| 68 | Asgher et al. | 2020 | 15 | 22 minutes | 12 | 8 | Prefrontal Cortex | N/A |
| 69 | Naseer et al. | 2016 | 7 | 440 seconds | 16 | 1.81 | Prefrontal Cortex | N/A |
| 70 | Hakimi et al. | 2020 | 20 | 10 minutes | 23 | 10 | Prefrontal Cortex | N/A |
| 71 | Erdoğan et al. | 2019 | 11 | 10 minutes | 48 | 3.91 | Frontal Cortex, Primary Motor Cortex and Somatosensory Motor | N/A |
| 72 | Hamid et al. | 2022 | 9 | 6 minutes | 12 | 1.81 | Left Hemisphere of M1 | N/A |
| 73 | Khan et al. | 2021 | 28 | 350 seconds | 48 | 3.9 | Frontal, Frontal-Central & Central Sulcus & Central & Temporal-Parietal Lobes | N/A |
| 74 | Ortega & Faisal | 2021 | 9 | 5 minutes | 24 | 12.5(80) | Bilateral Sensorimotor Cortex | N/A |
| 75 | Zhao | 2019 | 47 | | 24 | Not specified | Primary Motor & Prefrontal Region | N/A |
| 76 | Ghonchi et al. | 2015 | 29 | <30 minutes? | 36 | 10(128) | Not Specified | Open Access Dataset for EEG+NIRS Single-Trial Classification |
| 77 | Chiarelli et al. | 2018 | 15 | 10 minutes | 16 | 10 | Sensorimotor Regions | N/A |
| 78 | Cooney et al. | 2021 | 19 | 2 hours | 8 | 10(250) | Bihemispheric Motor Regions | N/A |
| 79 | Sun et al. | 2020 | 29 | 9 minutes | 36 | 12.5 | Not Specified | Open access dataset for EEG+NIRS single-trial classification |
| 80 | Kwak et al. | 2022 | 29 | 9 minutes | 36 | 12.5 | Not Specified | Open access dataset for EEG+NIRS single-trial classification |
| 81 | Khalil et al. | 2022 | 26 | 62 seconds | 36 | 10.4 | Frontal, motor cortex, parietal & occipital regions | Simultaneous acquisition of EEG and NIRS during cognitive tasks for an open access dataset |
| 82 | Xu et al. | 2020 | 47 | 8 minutes | 52 | 14.3 | Bilateral Temporal Lobe | N/A |
| 83 | Xu et al. | 2020 | 47 | 8 minutes | 44 | 14.3 | Bilateral Temporal Lobe | N/A |
| 84 | Ma et al. | 2020 | 84 | 2 minutes | 52 | 10 | bilateral frontal and temporal cortices | N/A |
| 85 | Wang et al. | 2021 | 96 | 150 minutes | 53 | 100 | Prefrontal Cortex | N/A |
| 86 | Chao et al. | 2021 | 32 | 15 minutes | 22 | 7.81 | Prefrontal Cortex | N/A |
| 87 | Chou et al. | 2021 | 67 | 160 seconds | 52 | 10 | Bilateral Frontotemporal Regions | N/A |
| 88 | Rosas-Romero et al. | 2019 | 5 | 30-180 minutes | 104-146 | 19.5 | Full Scalp Recording | N/A |
| 89 | Yang et al. | 2019 | 24 | 15 minutes | 48 | 8.138 | Prefrontal Cortex | N/A |
| 90 | Yang & Hong | 2021 | 24 | 5 minutes | 48 | 8.138 | Prefrontal Cortex | N/A |
| 91 | Ho et al. | 2022 | 140 | 30 minutes | 6 | 8 | Prefrontal Cortex | N/A |
| 92 | Behboodi et al. | 2019 | 10 | 9.5 minutes | 52 | 18.51 | Sensorimotor & Motor Areas | N/A |
| 93 | Sirpal et al. | 2021 | 40 | 75 minutes | 138 | 19.5 | Full Scalp Recording | N/A |
| 94 | Bandara et al. | 2019 | 20 | 17 minutes | 52 | 10 | Frontal Region | N/A |
| 96 | Qing et al. | 2020 | 8 | 18 minutes | 12 | 15.625 | Prefrontal Cortex | N/A |
| 97 | Ramirez et al. | 2022 | 5 | 14 minutes | 16 | Not specified | Left and Right Frontal Lobes | N/A |
| 98 | Hiwa | 2016 | 22 | 6.5 minutes | 24 | 10 | Left Hemisphere | N/A |
| 99 | Andreu-Perez et al. | 2021 | 30 | 7.5 minutes | 16 | Not specified | Prefrontal cortex | N/A |

**Table 2:** Description of the data collected in each paper that was considered.

| Citation No. | Author | Year | DL Architecture | General Task | Input | Output | Validation Metrics | Ground Truth | Results |
|---|---|---|---|---|---|---|---|---|---|
| | *Name of the Author* | *Published Year* | i.e. CNN, LSTM, etc. | i.e. Diagnosis | i.e Hbr, Hbo | *what is being output by the network* | i.e. LOSO, LOUO | *What was being used to assess the models* | *Put important numbers here.* |
| 19 | Mirbagheri et al. | 2020 | CNN | BCI | statistical features | Stress vs Relaxation | 5-Fold CV | Task labels for each trial | 88.52±0.77% Accuracy |
| 20 | Tanveer et al. | 2019 | MLP | Cortical Analysis | segmented time series | Extracted Features to Feed to Classifier | 10-Fold CV | Drowsiness detected via change in facial expression | 83.3±7.4% Accuracy with KNN Classifier |
| 21 | Dolmans et al. | 2021 | CNN+LSTM+MLP | BCI | segmented time series | Mental Workload Level | 5-Fold CV | Participant Reported Difficulty Ratings | 32% Accuracy |
| 23 | Saadati et al. | 2019 | DNN | BCI | segmented time series | n-back, WG, DSR & MI vs relaxation | LOOCV | | 89% Accuracy |
| 24 | Ortega & Faisal | 2021 | HEMCNN | BCI | statistical features | Left Hand Gripping or Right Hand | 5-Fold CV | Task labels for each trial | 78% Accuracy |
| 25 | Dargazany et al. | 2019 | MLP | BCI | raw fNIRS data | right hand, left hand, left leg, right leg & | | Task labels for each trial | 77-80% Accuracy |
| 26 | Nagasawa et al. | 2020 | WGAN | BCI | raw fNIRS data | Left hand ME, Right Hand ME, Bimanual | 10-Fold CV | N/A for data generation/Task labels for each trial | 73.3% Accuracy for Augmented SVM |
| 28 | Ghonchi et al. | 2020 | RCNN | BCI | 3-rank tensors of upsampled fNIRS time series | Mental Arithmetic or Rest/Motor Imagery | k-Fold CV | Task labels for each trial | 99.63% Accuracy |
| 29 | Trakoolwilaiwan et al. | 2017 | MLP/CNN | BCI | segmented time series | Left Hand MI Right Hand MI or Rest | 10-Fold CV | Task labels for each trial | 89.35% Accuracy for MLP & 92.68% Accuracy for CNN |
| 30 | Janani et al. | 2020 | MLP/CNN | BCI | Spectrograms+Sample-Point Images/ segmented time series | Left or Right Hand MI/ME | 5-Fold CV | Task labels for each trial | 80.49 ± 6.66% Accuracy (MI) & 85.66± 8.25% Accuracy (ME) |
| 31 | Yoo et al. | 2021 | LSTM | BCI | Time Series | English, Non-English, | 6-Fold CV | Task labels for each trial | 20.38± 4.63% Accuracy |
| 32 | Gao et al. | 2020 | CNN | Cortical Analysis | statistical features | Expert or Novice | LOUO+10-Fold CV | Reported skill level of participants | 91% Accuracy, 95% Sensitivity & 67% Specificity |
| 33 | Ortega et al. | 2021 | CNNATT | BCI | Upsampled fNIRS data | Discrete Force Profiles | | Simultaneously Recorded Grip Force | 55.2 FFAV% |
| 34 | Gao et al. | 2020 | CNNIRS | Preprocessing/ Augmentation | Simulated/Real HRF | Denoised HRF | | Simulated HRF Signals | 3.03 MSE |
| 35 | Ma et al. | 2021 | CNN | BCI | Time Series | Left Hand MI or Right Hand MI | LOSO CV | Task labels for each trial | 98.6% Accuracy with FCN & ResNet |
| 41 | Fernandez Rojas et al. | 2021 | LSTM | Diagnosis | raw HbO data | Low-cold, low-heat, high-cold or high- | 10-Fold CV | Task labels for each trial | 90.6% Accuracy 84.6% Sensitivity 90.4% Specificity |
| 42 | Wickramaratne Mahmud | 2021 | CNN | BCI | statistical features | MI MA or rest | 10-Fold CV | Task labels for each trial | 87.14±3.20% Accuracy |
| 44 | Sirpal et al. | 2019 | LSTM | Diagnosis | Time Series | Epileptic Siezure vs normal | 10-Fold CV | Labeled Siezure and Nonseizure Segments | 98.3±0.4% Accuracy, 89.7±0.5% Recall, 87.3±0.0.8% Precision |
| 46 | Xu et al. | 2019 | CNN+GRU | Diagnosis | segmented time series | ASD or TD | | Diagnosis of Subject | 92.2% accuracy, 85.0% sensitivity, and 99.4% specificity |
| 47 | Yang et al. | 2020 | CNN | Diagnosis | spatial maps from ΔHbO signals & spatio-temporal maps from | Cognitive Impairment or | 5-Fold CV | Diagnosis of Subject | 90.37±5.30% Accuracy, 86.98±7.25% Recall, 82.19±9.93% Precision for VFT task |
| 52 | Takagi et al. | 2020 | CNN | Cortical Analysis | Oxy, Deoxy & OD Images | Teeth Clenching or Relaxed | 5-Fold CV | Task labels for each trial | 90.3± 6.5% Accuracy, 88.1 ± 10.8% Recall, 92.4± 7.8% Specificity, 92.5± 7.1% Precision |
| 53 | Lu et al. | 2020 | LSTM+CNN | BCI | statistical features | Mental Arithmetic or Rest | 5-Fold CV | Task labels for each trial | 95.3% Accuracy |
| 54 | Saadati et al. | 2019 | CNN | BCI | Topographical Activity Maps | 0-back 2-back or 3-back task | 10-Fold CV | Task labels for each trial | 97±1% Accuracy |
| 56 | Benerradi et al. | 2019 | CNN | BCI | statistical features | Mental Workload Level | LOUO CV | Task labels for each trial | 49.53% Accuracy for 3 Classes & 72.77% Accuracy for 2 Classes |
| 60 | Liu et al. | 2021 | ESN/CAE | Preprocessing/ Augmentation | segmented time series | 0-back, 1-back, 2-back, or 3-back task | 10x10 CV | Task labels for each trial | 52.45 (ESN) & 47.21% (CAE) Accuracy for 4 classes |
| 61 | Lee et al. | 2018 | MLP | Preprocessing/ Augmentation | Time Series | Denoised Time Series | | Wavelet Denoised Methods | CNR of 0.63 |
| 62 | Kim et al. | 2022 | CNN | Preprocessing/ Augmentation | Time Series + HRF | Denoised Time Series | Ablation | Simulated HRF Signals | MSE of approx .004-.005 |
| 63 | Wickramaratne & Mahmud | 2021 | GAN+CNN | Preprocessing/ Augmentation | GASF/ kernel PCA GASF | GASF/Motor Task | LOOCV | N/A for data generation/Task labels for each trial | 96.67% Accuracy & 0.98 AUROC for CNN+110% Generated Data |
| 64 | Woo et al. | 2020 | DCGAN+CNN | Preprocessing/ Augmentation | HbO t-maps | ME or Rest | | Task labels for each trial | 92.42% for Unaugmented Data & 97.17% for Augmented Data |
| 65 | Hennrich et al. | 2015 | MLP | BCI | Not specified | Mental Arithmeetics, Word Generation, | 10-Fold CV | | 64.1% Accuracy |
| 66 | Kwon & Im | 2021 | CNN | BCI | segmented time series | MA vs Relaxation | LOSO CV | Task labels for each trial | 71.20± 8.74% Accuracy |
| 67 | Ho et al. | 2019 | DBN/CNN | BCI | statistical features | Mental Workload Level | n-Fold CV | Task labels for each trial | 84.26±2.58% Accuracy for DBN without PCA & 75.59±3.4% for DBN with PCA inputs & 72.77±1.92% Accuracy for CNN without PCA & 68.12±3.26% with PCA inputs |
| 68 | Asgher et al. | 2020 | LSTM | BCI | statistical features | Mental Workload Level | 10-Fold CV | Task labels for each trial | 89.31±3.95% Accuracy 87.51±3.90% Precision 86.76±4.38% Recall |
| 69 | Naseer et al. | 2016 | MLP | BCI | statistical features | MA or Rest | Ablation | Task labels for each trial | 96.3±0.3% Accuracy for MLP |

| # | Author | Year | Model | Application | Input | Task | Validation | Labels | Results |
|---|---|---|---|---|---|---|---|---|---|
| 69 | Naseer et al. | 2016 | MLP | BCI | statistical features | MA or Rest | Ablation | Task labels for each trial | 96.3±0.3% Accuracy for MLP |
| 70 | Hakimi et al. | 2020 | CNN | BCI | statistical features | Stress vs Relaxation | 5-Fold CV | Task labels for each trial | 98.69 ± 0.45% Accuracy for HRF feature set & 88.60 ± 1.15% Accuracy for fNIRS feature set |
| 71 | Erdoğan et al. | 2019 | MLP | BCI | statistical features | MI, ME or Rest | Ablation | Task labels for each trial | 96.3%±1.3% Accuracy for ME vs Rest, 95.8%±1.2% Accuracy for MI vs Rest & 80.1%±2.6% Accuracy for ME vs MI |
| 72 | Hamid et al. | 2022 | CNN/LSTM | BCI | Time Series | Motor Execution or Rest | 10-Fold CV | Task labels for each trial | 79.73% Accuracy with CNN, 77.21% Accuracy with LSTM & 78.97% Accuracy with Bi-LSTM |
| 73 | Khan et al. | 2021 | MLP | BCI | statistical features | Specific Finger Tapping or Rest | LOSO CV | Task labels for each trial | 60±2% Accuracy |
| 74 | Ortega & Faisal | 2021 | CNNATT | BCI | segmented time series | Force Profiles | 5-Fold CV | Simultaneously Recorded Grip Force | 55% FVAF |
| 75 | Zhao | 2019 | BiLSTM | BCI | statistical features | Goal execution vs Completion | | Task labels for each trial | 71.70% Accuracy |
| 76 | Ghonchi et al. | 2015 | LSTM/CNN | BCI | Upsampled fNIRS data | MI classes | 10x5-Fold CV | Task labels for each trial | 99.6% Accuracy |
| 77 | Chiarelli et al. | 2018 | MLP | BCI | segmented time series | Left Hand MI or Right Hand MI | 10-Fold CV | Task labels for each trial | 83.28±2.36% Accuracy |
| 78 | Cooney et al. | 2021 | CNN | BCI | filtered frequency bands of segmented time series | 1 of 4 action words + 1 of 4 word | nested 5-Fold CV | Task labels for each trial | 46.31% Accuracy for Overt Speech & 34.29% Accuracy for Imagined Speech |
| 79 | Sun et al. | 2020 | CNN | BCI | tensors of fused EEG&fNIRS data | Mental Arithmetic/Motor | 5-Fold CV | Task labels for each trial | 77.53% Accuracy for MI & 90.19% Accuracy for MA |
| 80 | Kwak et al. | 2022 | CNN+Attention | BCI | 3-rank tensors of upsampled fNIRS time series | Mental Arithmetic/Motor | Ablation | Task labels for each trial | 91.96±5.82% Accuracy for MA & 78.59±5.82% Accuracy for MI |
| 81 | Khalil et al. | 2022 | CNN | BCI | Time Series | Mental Workload Level | 10-Fold CV | Task labels for each trial | 94.52% Accuracy |
| 82 | Xu et al. | 2020 | LSTM | Diagnosis | Time Series | ASD or TD | 10-Fold CV | Diagnosis of Subject | 95.7±4.99% Accuracy |
| 83 | Xu et al. | 2020 | CNN+Attention | Diagnosis | segmented time series | ASD or TD | 10-Fold CV | Diagnosis of Subject | 93.3% Accuracy 90.6% Sensitivity & 97.5% Specificity |
| 84 | Ma et al. | 2020 | AttentionLSTM+CNN | Diagnosis | normalized HbO, HbR, & HbT matrices | BD or MDD | k-Fold CV | Diagnosis of Subject | 96.2% Accuracy |
| 85 | Wang et al. | 2021 | CNN | Diagnosis | Time Series/Statistical Features | Depressed or Non-Depressed | Ablation | Diagnosis of Subject | 83% Accuracy 79% Precision & 83% Recall (Manually Extracted Features), 72% Accuracy, 80% Precision & 75% Recall (Raw Data) |
| 86 | Chao et al. | 2021 | CFNN/RNN | Diagnosis | statistical features | Fear Stimulus or Rest | LOSO CV | Task labels for each trial | 99.94% Accuracy with CFNN & 99.94% with RNN |
| 87 | Chou et al. | 2021 | MLP | Diagnosis | statistical features | FES or Healthy | 7-Fold CV | Diagnosis of Subject | 79.7% Accuracy, 88.8% Specificity & 74.9% Specificity |
| 88 | Rosas-Romero et al. | 2019 | CNN | Diagnosis | 3-dimensioal tensors of HbO & HbR | Pre-ictal vs inter-ictal | 5-Fold CV | Labeled Siezure and Nonseizure Segments | 99.67±0.75% Accuracy for CNN |
| 89 | Yang et al. | 2019 | CNN | Diagnosis | Activation t-map / Channel correlation map | Cognitive Impairment or Healthy, | 6-Fold CV | Diagnosis of Subject | 90.62% Accuracy with t-maps & 85.58% with correlation maps |
| 90 | Yang & Hong | 2021 | CNN | Diagnosis | Connectivity Map | Mean STD & Variance of Δ HbO | | Diagnosis of Subject | 97.01% Accuracy |
| 91 | Ho et al. | 2022 | LSTM/LSTM+CNNN | Diagnosis | Segmented Time Series | asymptomatic, | 5-Fold CV | Diagnosis of Subject | 86.8% Accuracy for CNN-LSTM & 84.4% Accuracy for LSTM |
| 92 | Behboodi et al. | 2019 | MLP/CNN | Cortical Analysis | Time Series | RSFC Estimation | | Channels anatomically located over the motor and sensorimotor cortex | 0.89 AUC for MLP & 0.92 AUC for CNN |
| 93 | Sirpal et al. | 2021 | LSTM AE | Cortical Analysis | Full Spectrum EEG | ΔHbO concentration | k-Fold CV | Real fNIRS data | 6.52x10^-2 mean reconstruction error (Euclidean) |
| 94 | Bandara et al. | 2019 | CNN+LSTM | Cortical Analysis | segmented time series | High/Low Valence+ High/Low Arousal | 5-Fold CV | Emotional Valence & Arousal Scores of DEAP Dataset | 70.18% Accuracy for 1s windows & 77.29% Accuracy for 10s windows |
| 96 | Qing et al. | 2020 | CNN | Cortical Analysis | segmented time series | like/dislike, like/so-so, or dislike/so-so | 8-Fold CV | Subject Ratings for Each Trial | 84.3, 87.9, and 86.4% Accuracy for 15s, 30s, and 60s respectively |
| 97 | Ramirez et al. | 2022 | CNN | Cortical Analysis | fNIRS/fNIRS+EEG images | Discrete Preference Ratings | | Subject Ratings for Each Trial | 66.86% Accuracy with fNIRS & 91.83% Accuracy with fNIRS+EEG |
| 98 | Hiwa | 2016 | CNN | Cortical Analysis | time series | Male or Female | LOOCV | Gender of Subject | Approx. 60% Accuracy in 5 best channels |
| 99 | Andreu-Perez et al. | 2021 | DCAE/MLP | Cortical Analysis | statistical features | Novice, Intermediate, or | 10 repeated stratified k-fold | Reported skill level of participants | 91.43± 6.32% Accuracy for DCAE, 91.44± 9.97% Accuracy for MLP |

**Table 3:** The applications and key findings of each paper considered

## 2. DL applications in fNIRS

In this section of the review paper, we summarize and discuss the key findings of the literature search as well as how the techniques discussed above are currently being used. In order to properly understand the scope of this review, the literature search methodology will first be described. The primary method of searching for relevant articles was via the PubMed search engine. Papers published between 2015 and May 2022 in which deep learning was used in conjunction with fNIRS data were considered for this review. Due to very few fNIRS papers being published using deep learning prior to 2015 the search was restricted to this time frame. Using the terms "deep learning" and "fNIRS" resulted in 35 results, 29 of which were relevant, while the terms "deep learning" and "NIRS" only produced 12 articles, none of which were both relevant and absent from the previous search. When searching the terms "neural network" and "fNIRS", 146 results were found, with many using the term neural network to refer to the neuroscience phenomenon being studied through the use of deep learning. As a result only 3 additional relevant papers were found via this search. In addition to this, using the same search terms in google scholar produced tens of thousands of results, many of which were not relevant. Due to the impracticality of a comprehensive search of these results, the search was truncated after multiple pages of results yielded no relevant articles. From this google scholar search, an additional 25 articles were found, yielding a total of 63 articles which were considered in this review.

In recent years, the use of DL techniques in fNIRS studies has increased, and due to the versatility of fNIRS, DL has been applied to many different applications of fNIRS. While some of the studies which used DL used it for feature extraction or data augmentation, in most of the papers considered, DL was used as a classifier. As a result, those studies in which DL was used as a classifier are further subdivided into categories based on the application of fNIRS in the study. A comprehensive summary of the applications and DL architecture employed is Table 3, while the details of the experimental setups for the fNIRS studies are summarized in Table 2. Because fNIRS is of interest in studies on BCI, many of the studies found used DL classifiers for the classification of tasks for BCI applications. Other studies have used DL techniques as diagnostic tools, to detect various physiological and mental pathologies based on cortical activity. Finally, some studies, such as those using DL techniques to assess skill level and functional connectivity, were not common enough to be placed into a category of their own, however these papers all focused on the analyses of cortical activity using DL techniques, and as such are grouped together. **Figure 3** is provided for visualization of the number of papers collected from each category for the given year. While some of the papers mentioned may fall within multiple categories, additional context from the paper such as the primary focus of the paper assisted in determining how the paper was categorized. This did not stop relevant papers from being discussed in more than one subcategory.

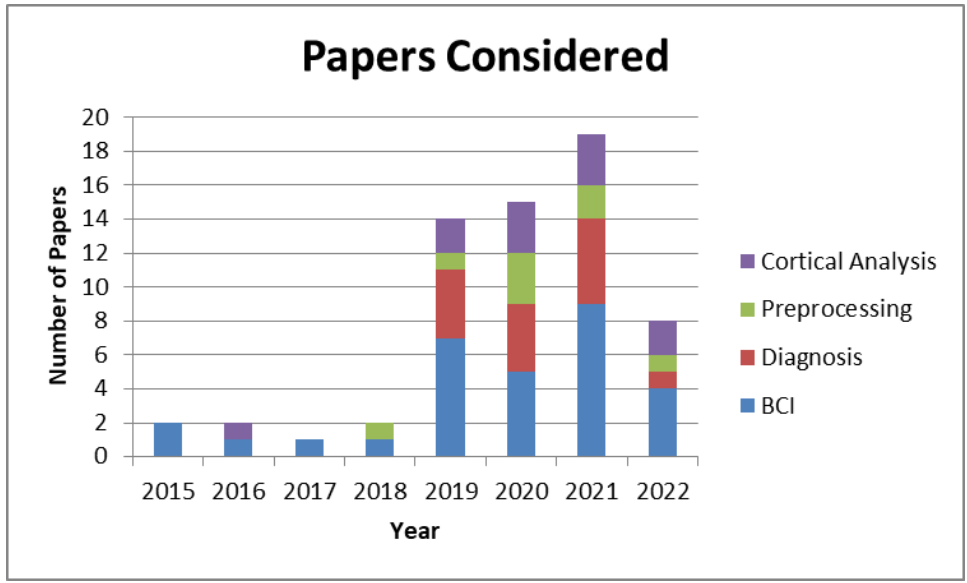

**Figure 3:** Distribution of the 63 papers reviewed in this article by year and color coded by application field.

**2.1 Preprocessing**

Like most noninvasive neuroimaging modalities, raw fNIRS signals typically contain confounding physiological signals and other noise which originates from outside of the cerebral cortex such as the hemodynamics of the scalp and changes in blood pressure and heart rate [55]. Most studies use some method of preprocessing to try to address this. While many of the papers presented here use band pass filtering or butterworth filtering, various other methods are employed throughout the literature. ICA denoising [12], wavelet filtering [29] and correlation based signal improvement (CBSI) filtering [56] are all methods used to remove some of the undesired physiological trends in fNIRS signals. Another recommended method is short separation regression, a method in which signals with only confounding physiological signals are simultaneously collected alongside fNIRS signals and the trends in these signals are removed from the fNIRS data [57]. Other algorithms such as Savitsky-Golay filtering [58] and temporal derivative distribution repair (TDDR) [59] are methods used to correct motion artifacts and baseline drifts in fNIRS signals. Many of these techniques are commonly used in fNIRS studies in order to improve the quality of the signal as well as ensure that the signal being assessed originates from the brain. Other recommended techniques involve prewhitening the data or decorrelating via PCA to remove any correlation between fNIRS signals prior to analysis [57], further facilitating the analysis of signals originating in a region of interest.

**2.2 Feature Extraction and Data Augmentation**

One of the most notable problems with fNIRS data is the extensive manual feature extraction and artifact removal typically required in order for data to be analyzed, preventing many fNIRS studies from being applied in real time. As a result, more effective methods of preprocessing fNIRS data are being explored. One of the most promising benefits of DL is the ability to quickly and automatically learn and extract relevant features in fNIRS data. Some studies have already explored this benefit of DL. Tanveer et al. [20] reported the use of a DNN to extract the features which were fed to a K-nearest neighbors (KNN) classifier in order to detect the drowsiness of subjects during a virtual driving task. Using the features extracted from the DNN, the KNN was able to achieve a classification accuracy of 83.3%. Despite feature extraction typically being computationally expensive, even taking hours with a powerful GPU, the DNN exhibited a mean computation time of 0.024 seconds for 10s time windows, a speed which would allow for feature extraction of a 30 minute signal to take less than 5 seconds with a NVidia 1060 GTX GPU. On top of computational speed, the ability of DL to automatically learn and extract features may reduce bias and errors during feature extraction, allowing for an increase in classification accuracy. One study by Liu et al. [60] used an echo state network auto-encoder (ESN AE) to extract the features that were fed to a multilayer perceptron (MLP), achieving a four-class classification accuracy of 52.45%, outperforming the accuracy of the convolutional auto-encoder (CAE)+CNN and manually extracted features fed to an MLP which achieved accuracies of 47.21% and 37.94% respectively.

While there have been other papers which are interested in using DL to extract features to feed into another classifier, there have also been papers which take raw fNIRS data and use the same neural network for feature extraction and classification. Despite end-to-end neural networks being seen as a more ideal solution than manual feature extraction, difficulties with low generalizability make them less commonly used. One study by Dargazany et al. [25] used an MLP (with 2 hidden layers) with raw EEG, body motion and fNIRS data to achieve a reported classification accuracy of 78-80% on a motor task with four classes, despite no de-noising or preprocessing of data being done. Another study by Rojas et al. [41] used raw fNIRS data as the input for an LSTM network, achieving a classification accuracy of 90.6%. In order to assess generalizability, a 10-fold cross validation was used, with the classifier achieving an accuracy of 93.1%. Both studies have provided evidence otwards the claim that DL techniques are capable of removing the need for manually extracted features from fNIRS data, further progressing towards the real-time end-to-end BCI. Despite the fact that the previously mentioned studies were able to achieve high accuracies without de-noising or motion artifact removal, some studies are performed when a subject's head is in motion. In such studies, fNIRS data can be heavily compromised by specific artefacts in the raw data that could bias any classification task. While there are many algorithms which are used to try to remove motion artifacts, Lee et al attempted to use a CNN to recognize and remove motion artifacts without relying on the parameters which must be defined in order to use many of the popular motion artifact removal algorithms [61]. In this study, the raw fNIRS time series and the estimated canonical response were used as inputs to the network and the resulting CNR of the DL output was 0.63, which outperformed wavelet denoising which achieved a mean CNR of 0.36. Another study done by Gao et al. [34] used subjects who were performing a precision cutting surgical task based on the Fundamentals of Laparoscopic Surgery (FLS) program which required a large range of motion. With a de-noising auto-encoder (DAE) and the process seen in **Figure 4**, 93% motion artifact removal in simulated data and 100% artifact removal in real data were reported, outperforming all comparable artifact removal techniques, including wavelet filtering and principal component analysis. Another study to look at DL for the removal of motion artifacts, Kim et al. [62] compared a CNN to the performance of wavelet denoising and an autoregressive denoising method. Using simulated data in a manner similar to Gao et al. to determine the ground truth, Kim et al. found that the CNN resulted in a mean square error (MSE) of approximately 0.004-0.005, while the next best method, the combination of wavelet and autoregressive denoising resulted in an MSE of approximately 0.009. Despite these studies all using different metrics to measure the effectiveness of the network, there is clearly an interest in finding an effective method of removing motion artifacts from fNIRS data.

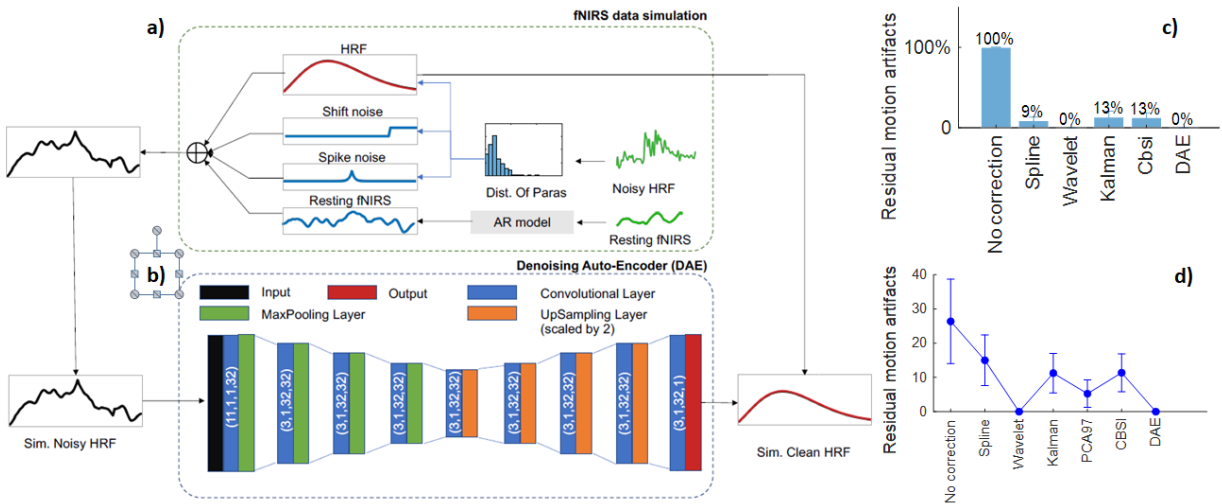

**Figure 4:** The illustration of the fNIRS data simulation process and the designed denoising auto-encoder (DAE) model. (a) The green lines are the experimental fNIRS data, including noisy HRF and resting fNIRS data, while the blue and red lines are simulated ones. (b) DAE model: The input data of the DAE model is the simulated noisy HRF, and the output is the corresponding clean HRF without noise. The DAE model incorporates nine convolutional layers, followed by max-pooling layers in the first four layers and upsampling layers in the next four layers, with one convolutional layer before the output. The parameters are labeled in parentheses for each convolutional layer, in the order of kernel size, stride, input channel size and kernel number. (c-d) number of residual motion artifacts for the simulated and experimental data sets respectively. Adapted from Gao et al. [34]

It is clear that DL techniques have shown promise for the ability to process and extract features from fNIRS data, but due to a lack of large variety of open-source fNIRS datasets, there has been a recent interest in using deep learning to generate more fNIRS data for training models. Data augmentation is a technique in which new data is generated in order to reduce the need for large labeled datasets for many machine learning and deep learning algorithms which require lots of labeled data. In order to be useful, this generated data must not be identical to any of the training data, but must also be realistic, i.e. it must remain within the distribution of the original dataset [63]. While this is a challenge, some deep learning techniques such as Generative Adversarial Networks (GANs), have been used to accomplish this. Wickramaratne and Mahmud [63] have used a GAN in order to augment their fNIRS dataset in order to increase the classification accuracy of finger and foot tapping tasks. Without augmented data, a classification of 70.4% was achieved with an SVM classifier, and a CNN classifier achieved an accuracy of 80% when trained only on real data. Using a GAN to generate training data, the accuracy of the CNN classifier increased to 96.67% when trained on real data as well as 110% generated data. Similarly, Woo et al. [64] used a GAN to produce activation t-maps, which when used to augment the training dataset of a CNN, increased the classification accuracy of a finger tapping task from 92% to 97%, showing that a network which is already performing well may benefit from data augmentation. While the previous studies have used GANs to generate an image representation of fNIRS data, only one study directly used a GAN to augment the dataset with raw time series fNIRS data. Nagasawa et al. [26] used a GAN to generate fNIRS time series data in order to increase the classification accuracy of motor tasks as shown in **Figure 5**. They reported that when augmenting the 16 original datasets with 100 generated datasets, the accuracy of the SVM classifier increased from around 0.4 to 0.733 while the accuracy of the neural network classifier increased from around 0.4 to 0.746. While still a relatively recent development in fNIRS studies, generated datasets using GANS have demonstrated the ability to increase the classification accuracy of commonly used classifiers such as CNN or SVM classifiers, once again demonstrating the versatility of DL techniques in fNIRS research.

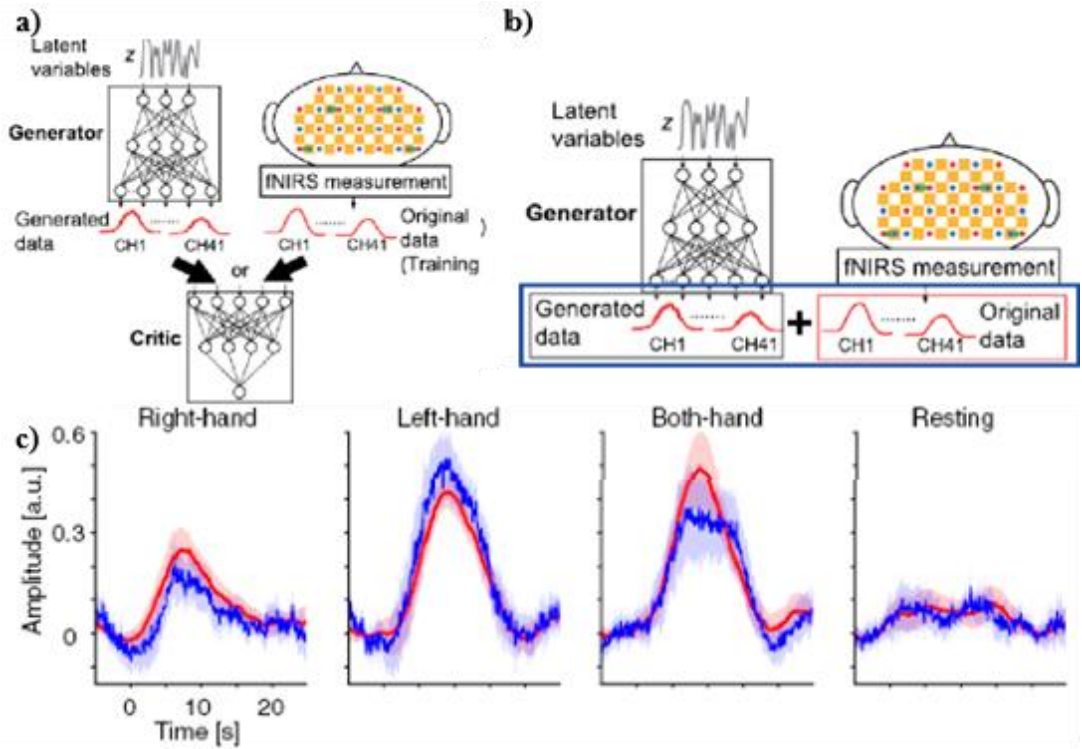

**Figure 5:** Framework of GAN and data augmentation. (a) The generator creates the data from the random variables z, and the critic evaluates the generated and original (measured) data. (b) After the training process, the data generated by the generator (referred to as generated data) are combined with the original fNIRS data as augmented data. (c) Trial-averaged waveforms for the 4 tasks considered in a cross-validation hold. The red lines denote measured original data and the blue lines denote generated data using WGANs. The shaded area represents 95% confidence intervals. Adapted from Nagasawa et al. [26]

## 2.3 Brain Computer Interface

One of the most promising applications for fNIRS research is BCI. Many studies on BCI use traditional machine learning or DL techniques in order to identify a certain task based on cortical activation. In 2015 Hennrich et al. [65] used a deep neural network (DNN) to classify when subjects were performing mental arithmetic, word generation, mental rotation of an object and relaxation. The reported accuracy of the DNN with 2 hidden layers was 64.1% which is comparable to the 65.7% accuracy of the shrinkage LDA despite the LDA using custom-built features while the input for the DNN was de-noised and normalized fNIRS data. In 2021, Kwon and Im [66] used similar inputs for their CNN, de-noised and baseline corrected $\Delta$HbR and $\Delta$HbO$_2$ data, in order to classify mental arithmetic from an idle fixation task. This CNN achieved a classification accuracy of 71.20% which surpassed the 65.74% accuracy of the shrinkage LDA classifier that used feature vectors as inputs. Wickramaratne and Mahmud [42] also used a CNN in order to classify mental arithmetic from an idle fixation task in 2021. Like Kwon and Im, a shrinkage LDA with feature vectors was used as the input. Unlike the previous study however, the inputs for the CNN were Gramian Angular Summation Fields (GASF), which are a type of image that is constructed from time series data that maintains some temporal correlation between points. With this CNN and GASF inputs, a classification accuracy of 87.14% was achieved, once again outperforming the shrinkage LDA which achieved an accuracy of 66.08%. Similarly, Ho et al. [67] also used a two-dimensional representation of fNIRS data to try and discriminate between differing levels of mental workload. In this study Ho et al. found that using a CNN with spectrograms generated from fNIRS data achieved a classification accuracy of 82.77%. This was outperformed by a deep belief network, which is a type of network similar to a MLP. The deep belief network, using manually extracted features from the HbR and $HbO_2$ signals achieved an accuracy of 84.26%. In a similar mental workload task, Asgher et al. [68] found that using an LSTM with similarly extracted features from HbR and $HbO_2$ time signals resulted in a mental workload classification accuracy of 89.31%. While many of the studies presented compared the performance of deep learning techniques to that of traditional machine learning or other algorithms, Naseer et al. [69] compared the classification accuracy of an MLP to that of a kNN, Naive Bayes, SVM, LDA & QDA algorithm. On a 2-class mental workload task, the MLP achieved an accuracy of 96% while the QDA, Naive Bayes and SVM classifiers all achieved similar accuracies of about 95% and the LDA and kNN algorithms performed much worse, with accuracies of 80% and 65% respectively. All classifiers in this study used manually extracted features such as skewness and kurtosis of the fNIRS as inputs into the algorithms. Hakimi et al. [70] also used manually extracted features from fNIRS time series in order to perform 2-class classification between stress and relaxation states and with a CNN, achieved an accuracy of 98.69%, further reinforcing that DL can give very high classification accuracies with mental workload tasks.

While classification of mental tasks such as arithmetic or word generation is commonly used, many forms of BCI are designed to help those who have restricted or reduced motor function. Because of this, another popular type of experiment found in BCI studies involves executing motor tasks. Trakoolwilwaiwan et al. [29] was able to achieve an accuracy of 92.68% with a CNN in a 3-class test to distinguish the finger tapping of the left hand, right hand and both hands at rest despite the CNN being used as both a feature extractor and classifier. The CNN outperformed the SVM and ANN classifiers, which reported accuracies of 86.19% and 89.35% respectively which were given the extracted feature inputs commonly used in BCI fNIRS studies (signal mean, variance, kurtosis, skewness, peak and slope). Since much of the interest in BCI applications involves aiding those who lack the ability to move, some studies focus not on the cortical activations of movement, but rather on the cortical activations of imagining movement. In one of these motor imagery studies, Janani et al. [30] had subjects perform a hand clenching or foot tapping task, and shortly after, imagine themselves performing the same task.

Two different types of input images were used in order to see which method a CNN classifier would more effectively extract features from. The first type of input image turned all of the data points within a 20 second window into an MxN matrix, where M is the number of data points and N is the number of channels. The second type of input used a short-time Fourier transform to turn the one-dimensional data into two-dimensional time-frequency maps of each channel that were stacked on top of each other to form an input image.One study by Erdoĝan et al. [71] similarly performed classification between motor imagery vs motor execution using an MLP with a classification accuracy of 96.3% between finger tapping and rest and 80.1% accuracy between finger tapping and imagined finger tapping when using manually extracted features from fNIRS data. Hamid et al. [72] attempted to distinguish between a treadmill walking task and rest using bandpass filtered fNIRS data and an LSTM. The LSTM achieved an accuracy of 78.97%. When compared to traditional classifiers which had statistical features manually extracted, the kNN achieved the next best performance of 68.38% accuracy. The SVM and LDA were also outperformed by DL, achieving accuracies of 66.63% and 65.96% respectively, despite using manually extracted features as inputs. This demonstrates the ability for DL to perform well on fNIRS data without requiring the extensive processing or feature extraction typically used in conjunction with other classifiers for fNIRS data. For many of these BCI motor tasks, being used for prosthetics would be more applicable. In these situations, more fine motor control using fNIRS would need to be assessed. Khan et al. [73] addressed this by performing 6-class classification between rest and each finger on the right hand of the subjects, achieving an accuracy of 60%. Ortega and Faisal [24] attempted to distinguish between a left and right hand gripping task by using a PCA to reduce dimensionality of the denoised time series data before feeding the segmented time series into a CNN based architecture. The resulting accuracy of this study was 77%. To further investigate this, Ortega et al. [33] used a CNN with attention and simultaneously recorded EEG signals in order to try to reconstruct the grip force of each hand during the task. This resulted in an average fraction of variance accounted for of 55% when reconstructing the discrete grip force profiles, demonstrating that not only can the DL techniques distinguish between which hand was performing a motor task, but also displays progress towards using fNIRS and EEG signals to determine the amount of force exerted during that motor task. Ortega and Faisal [74] then attempted to use this architecture to determine force onset and which hand was providing more force. This resulted in a force onset detection of 85% but only a hand disentanglement accuracy of 53%, showing that there is still progress to be made towards the complex decoding and reconstruction of motor activities.

In the motor imagery tasks, the first input image method achieved an accuracy of 77.58% using $HbO_2$ data, while the second method achieved an accuracy of 80.49% using $HbO_2$ data. It could be noted that HbR and HbT were also tested but for both methods, $HbO_2$ showed consistently higher results. While $HbO_2$ is commonly used in fNIRS studies due to higher SNR, Yucel et al. and Herold et al. [55, 57] reported that trends found in $HbO_2$ signals but not in HbR signals may be due to higher sensitivity of $HbO_2$ to systemic signals not originating in the brain. For this reason, it is generally recommended that both HbR and $HbO_2$ signals are assessed in fNIRS studies. Other traditional classifiers were also used and the closest results were achieved by the meta-cognitive radial basis function network, which achieved a classification accuracy of 80.83%. Another study that focused on motor imagery, Ma et al. [35], used a type of time series data that included a one-hot label as the input for the neural networks. Of the DL techniques used, the fully connected network (FCN) and residual network (ResNet) achieved the highest average accuracy of 98.6%. Of the traditional machine learning techniques tested, the SVM had the highest classification accuracy of 94.7%. Interestingly enough, for the DL networks, the mean classification accuracy achieved with $HbO_2$+HbR+HbT data, 98.3%, was the same as the accuracy when only HbR+HbT data was input, which was attributed to the feature extraction capabilities of the DL techniques. This study also evaluated the accuracy of individual channels, with single-channel classification accuracy ranging from 61.0% to 80.1% with the three highest accuracy channels being found in the somatosensory motor cortex and primary motor cortex.

While many studies have found considerable success in distinguishing between certain tasks using cortical activations, most studies use simple tasks in controlled environments, and focus on distinguishing tasks from each other and resting state. For a practical BCI, more complex tasks and classification methods will need to be considered. Zhao et al. [75] addressed this by having participants perform a task in which they would pick up a table tennis ball with chopsticks and lift it about 20 cm in the air, using their non-dominant hand. An LSTM with was used to try to determine when the task was completed, and $\Delta HbO_2$ data was used as the input, resulting in an accuracy of 71.70%, outperforming the 66.6% accuracy of the SVM that was given mean, variance, kurtosis and skew features of the fNIRS data as inputs.

One commonly used technology for BCI studies is EEG, due to the high temporal resolution as well as portability and noninvasiveness. Because it has a high temporal resolution, but low spatial resolution, it is common to combine EEG measurements with fNIRS, due to both being portable and noninvasive. In a motor imagery study, Ghonchi et al. [76] used fNIRS to augment the EEG data being collected. Three types of DL networks were used as classifiers, a CNN for its capability to extract special features, an LSTM for its ability to extract temporal features, and a recurrent CNN (RCNN) for its ability to extract both temporal and spatial features. The RCNN achieved the highest classification accuracy of 99.6% when both EEG and fNIRS data was used. Interestingly, the accuracy of both the CNN and LSTM increased when fNIRS data was added, jumping from 85% and 81% to 98.2% and 95.8% respectively. This indicates that despite EEG data having higher temporal resolution, fNIRS data still contributes both spatial and temporal information. A 2016 study by Chiarelli et al. [77] also found an increase in classification accuracy when combining EEG and fNIRS data. When performing 2-class classification on a motor imagery task with an MLP, the average accuracies of EEG and fNIRS data alone were 73.38% and 71.92% respectively, but when using both modalities, accuracy increased to 83.28%, further reinforcing that the simultaneous acquisition of EEG and fNIRS can provide more relevant information than either modality on their own. Cooney et al. [78] found that when combining fNIRS and EEG data, they were able to distinguish between multiple combinations of overt speech with a CNN classifier, achieving an accuracy of 46.31%. When tested on imagined speech, the classifier achieved an accuracy of 34.29%, which is also higher than the random chance value of 6.25% for 16 possible combinations, which shows promise in the use of EEG and fNIRS for assisting patients who may be unable to verbally communicate. Sun et al. [79] used a CNN to try to distinguish between rest mental arithmetic and motor imagery tasks using both EEG and fNIRS data by generating tensors of the fused EEG and fNIRS data. For the motor imagery tasks, this resulted in an accuracy of 77.53% and for the mental arithmetic tasks, this resulted in an accuracy of 91.83%. Using the same dataset, Kwak et al. [80] applied a branched CNN architecture which used the fNIRS data to generate spatial feature maps which were then fed to the EEG maps to try to obtain higher spatial resolution than ordinary EEG and higher temporal resolution than fNIRS. The resulting classification accuracies were 78.97% for motor imagery and 91.96% for mental arithmetic tasks, which are only slight improvements over the tensor fusion methods of Sun et al. Khalil et al. [81] also used a fusion of fNIRS and EEG data to distinguish between rest and a mental workload tasks. With a CNN, an accuracy of 68.94% was achieved when trained on data from 5 of the 26 participants. When training on 16 participants then performing transfer learning to an additional 5 participants, accuracy increased to 94.52%, exemplifying not only how important larger datasets are for deep learning, but how transfer learning can be used to address this. It may be of interest to explore the use of transfer learning from other fNIRS datasets, which use different hardware systems, different tasks or different fNIRS channel arrangements, since many studies rely on collecting fNIRS data specific to their own applications. As a result, it would be important to see if transfer learning can help extract meaningful features from fNIRS data independently of the part of the brain being recorded or the hardware being used. While many studies have only recently begun looking to use fNIRS for real-time BCI applications, current studies in the field have found use in DL techniques for increased classification accuracy and automatic feature extraction.

**2.4 Diagnostic Tools**

One promising use of DL techniques with fNIRS is in clinical applications, most notably as a diagnostic tool. In 2019, Xu et al. [46] recorded resting state fNIRS data from the bilateral frontal gyrus and bilateral temporal lobe of children. Using a single channel in the left temporal lobe, a CGRNN classifier was able to successfully classify autism spectrum disorder (ASD) in children with 92.2% accuracy, 85.0% sensitivity and 99.4% specificity for 7 seconds of resting-state HbR data. As shown in **Figure 6**, multiple $HbO_2$ and HbR channels showed statistically significant differences between the group with ASD and the control group. In 2020, Xu et al. [82] used a CNN classifier with attention layers and achieved an accuracy sensitivity and specificity of 93.3%, 90.6% and 97.5% respectively using similar resting state data to classify between ASD and typically developing (TD) subjects. Xu et al. [83] further explored using fNIRS data to detect autism is subjects and found that by using a CNN+LSTM classifier and $HbO_2$ data, they were able to achieve a single-channel accuracy, sensitivity and specificity of 95.7%, 97.1% and 94.3% respectively for the detection of ASD. Aside from autism, the use of fNIRS to diagnose other psychological disorders has been studied. With an average classification accuracy of 96.2%, Ma et al. [84] was able to distinguish bipolar depression from major depressive disorder in adults during a verbal fluency task by using an LSTM. Wang et al. [85] managed to distinguish between healthy subjects and those diagnosed with major depressive disorder with an accuracy of 83.3%. This was accomplished using long recording times of 150 minutes each from a relatively large sample size of 96 subjects. As can be seen in Table 2, this is a larger sample size than any of the other fNIRS studies presented, which lends to confidence in the generalizability of this neural network to new subjects. In 2021, Chao et al. [86] used a cascade forward neural network (a network similar to an MLP) to perform and achieved an average classification accuracy of 99.94% between depressed and healthy subjects when a fear stimulus was presented across 32 subjects. Chou et al. [87] used an MLP network to classify between subjects with first episode schizophrenia and healthy subjects, achieving a classification accuracy of 79.7% with only about 160 seconds of recorded data from each subject.

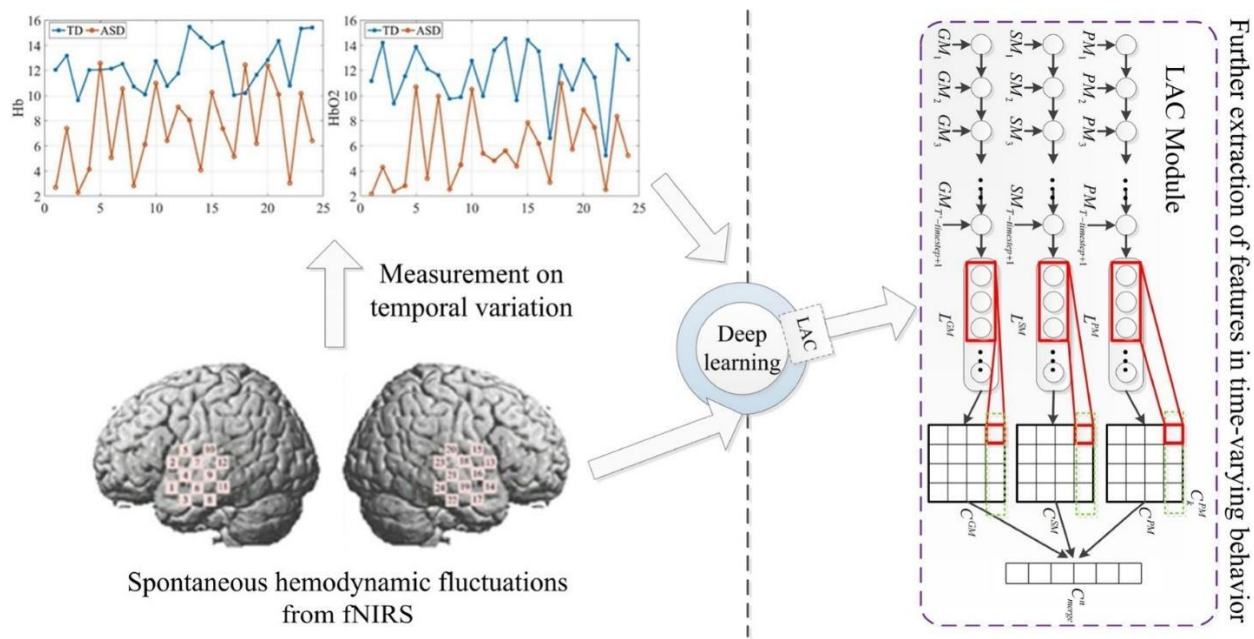

**Figure 6:** A deep learning model combining LSTM and CNN (LAC) was used to accurately identify autism spectrum disorder (ASD) from a typically developing (TD) subject based on time-varying behavior of spontaneous hemodynamic fluctuations from fNIRS. Adapted from Xu et al. [83]

EEG technology is commonly used in clinical settings, however, some studies have used EEG alongside fNIRS for diagnostic studies. Sirpal et al. [44] used an LSTM architecture and fNIRS data to detect seizures with 97.0% accuracy, and with EEG data, achieved 97.6% accuracy, but with combined EEG and fNIRS data, accuracy increased to 98.3%, once again displaying how hybrid EEG-fNIRS recordings can increase classification accuracy, even when accuracy is already high. Rosas-Romero et al. [88] also combined fNIRS and EEG signals to detect epilepsy. Despite only having recordings from 5 subjects, a CNN was able to detect pre-ictal segments fNIRS and EEG data with an average accuracy of 99.67% with 5-Fold cross validation.

Another use for fNIRS is to help detect mild cognitive impairment (MCI), the prodromal stage of Alzheimer's disease. In 2019, Yang et al. [89] employed three different strategies using CNNs to detect MCI. Using an N-back, Stroop and verbal fluency (VFT) task, a CNN trained on concentrations changes of $HbO_2$ achieved accuracies ranging from 64.21% in the N-back task to 78.94% in the VFT task. When using activation maps as the inputs for the CNN, the accuracy ranged from 71.59% in the VFT task to 90.62% in the N-back task. The final strategy employed, using correlation maps as inputs showed lower accuracies than the activation maps, with the highest accuracy being 85.58% for the N-back task. In 2020, Yang et al. [47] used temporal feature maps as inputs for the CNN classifier, resulting in average accuracies of 89.46%, 87.80% and 90.37% with the N-back, Stroop and VFT task respectively. In another study done in 2021 by Yang and Hong [90], pre-trained networks were used to distinguish between subjects with MCI and the healthy control group. Using resting state fNIRS data, the network with the highest accuracy, VGG19, achieved an accuracy of 97.01% when connectivity maps were used as the input, outperforming the conventional machine learning techniques, with LDA classifier reporting the highest accuracy of 67.00%. Ho et al. [91] attempted to use fNIRS and DL techniques to distinguish not only between healthy and prodromal Alzheimer's afflicted subjects, but also subjects with asymptomatic Alzheimer's disease and dementia due to Alzheimer's disease. Not only did this study try to distinguish between different stages of Alzheimer's disease, the study used a notably large sample size of 140 participants, which was larger than any other study reported in this review as seen in Table 2. The 86.8% accuracy of the CNN-LSTM network when 5-Fold cross validated not only shows the ability of the network to distinguish between a wide range of subjects with and without Alzheimer's disease, it also shows the ability of this network to distinguish between different stages of Alzheimer's disease, making this a very promising tool for clinical use.

Yet another clinical application for fNIRS measurements with DL techniques was explored by Rojas et al. [41], where raw fNIRS data and an LSTM were used to distinguish between high and low levels of pain as well as whether the pain was caused by a hot or cold stimulus with an achieved accuracy of 90.6%. Being able to assess the intensity of pain as well as the cause of it could be exceptionally useful in instances where patients are unable to communicate, such as with non-verbal patients. This further displays the usefulness of fNIRS with DL techniques as a robust and accurate diagnostic tool.

**2.5 Analysis of Cortical Activations**

Outside of BCI and diagnostic tools, fNIRS data still has many uses. Understanding the functional connectivity of the brain is an essential part of understanding the mechanisms behind numerous neurological phenomena. As a result, there is an interest in using neuroimaging to understand the functional connectivity of the brain. Behboodi et al. [92] used fNIRS to record the resting state functional connectivity (RSFC) of the sensorimotor and motor regions of the brain. In this study, four methods were used, seed based, independent component analysis (ICA), ANN and CNN. Unlike the first two methods, very limited preprocessing was used for the ANN and CNN, with both using filtered $HbO_2$ data to form the connectivity maps. Each connectivity map was then compared to the expected activation based on the physiological location of each detector, which was used as the ground truth, using an ROC curve, as shown in **Figure 7**, with the CNN achieving an AUC of 0.92, which outperformed the ANN, ICA and seed-based methods with each reporting an AUC of 0.89, 0.88 and 0.79 respectively. Another study which investigated RSFC, Sirpal et al. [93], collected EEG and fNIRS data, and attempted to use the EEG data and an LSTM to recreate the fNIRS signals. Using only the gamma bands of the EEG signal was found to have the lowest reconstruction error, below 0.25. The reconstructed fNIRS signals were further validated by comparing the functional connectivity of the signals constructed using only gamma bands and those using the full spectrum EEG signals with the functional connectivity of the experimental fNIRS data. Despite the signals formed using gamma bands only having a lower reconstruction error, the root mean square error of the full spectrum EEG signals was consistently lower.

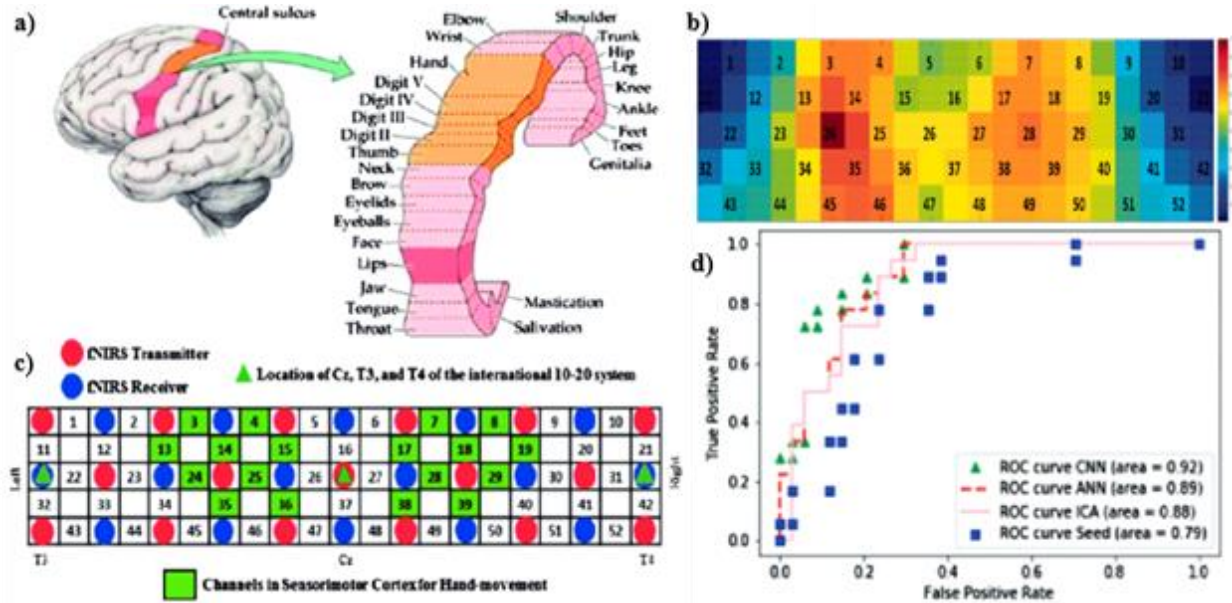

**Figure 7:** (a) The cortical areas of hand-movement in sensorimotor and motor cortices (the picture is adopted from http://neurones.co.uk). (b) Anatomical areas of sensorimotor and motor areas which are related to the hand-movement monitored by fNIRS are highlighted in green. (c) Group resting-state functional connectivity (RSFC) map derived from CNN-based resting-state connectivity detection. (d) ROC curves on different methods. Adapted from Behboodi et al. [92]

Other types of fNIRS studies have also been done which utilized DL. Some studies have analyzed the cortical activations of subjects to predict emotions. Bandara et al. [94] used music videos from the DEAP database [95] to classify the emotional valence and arousal of subjects using a CNN+LSTM architecture and fNIRS data recorded from the prefrontal cortex. Using the subjects' self-assessments as ground-truth, a classification accuracy of 77% was reported. Another study collecting fNIRS signals from the prefrontal cortex, Qing et al [96] used a CNN to determine the preference levels of subjects towards various Pepsi and Coca-Cola ads, achieving an average three-class classification accuracy of 87.9% for 30 second videos. 15 and 60 second videos showed similar accuracies of 84.3% and 86.4% respectively. Similarly, Ramirez et al. [97] attempted to decode consumer preference towards 14 different products. With a CNN, Ramirez et al. was able to distinguish between a strong like and strong dislike of the presented product with an accuracy of 68.6% with fNIRS data, 77.98% with just EEG data and 91.83% with combined fNIRS and EEG data. Hiwa et al. [98] studied the use of fNIRS and CNNs for the identification of a subject's gender, achieving an accuracy of approximately 60% when using the filtered fNIRS data from only 5 channels.

While most studies have focused on using cortical activations to classify when a specified task is being completed, a few studies have begun looking into predicting the skill level of the subject at a given task. Andreu-Perez et al. [99] classified the expertise of subjects watching 30 second clips of the video game League of Legends by using fNIRS data and facial expressions. The fully connected deep neural network (FCDNN) and deep classifier auto encoder (DCAE) were compared against many traditional machine learning techniques, including SVM and kNN in a three-class test to determine the skill level of the subject watching. Using only fNIRS data, the DL classifiers had the two highest accuracies of 89.84% and 90.70% for the FCDNN and DCAE respectively while the most accurate machine learning technique, SVM, only achieved an accuracy of around 58.23%. When the predicted emotion scores were also included, the accuracy of the FCDNN and DCAE improved to 91.44% and 91.43% respectively. Most of the machine learning techniques also saw minor or no improvements, with the SVM still achieving an accuracy of 58.69%. The XGBoost classifier saw a large increase in accuracy when emotion scores were added, increasing from 50.79% to 71.55%, which was still about 20% lower than the accuracies of the DL techniques used. Another study by Gao et al. [32] looked to predict the surgical skill level of medical school students who were being assessed on tasks based on the Fundamentals of Laparoscopic Surgery (FLS) protocols. In this study, subjects would perform an FLS precision cutting task while fNIRS data was recorded from the prefrontal cortex. The subjects would be assessed and scored in accordance with FLS procedures. A Brain-NET architecture was used to predict the FLS scores of each participant based on the features extracted from the recorded fNIRS data. The Brain-NET model reported an ROC AUC of 0.91, outperforming the kernel partial least squares (KLPS), random forest and support vector regression (SVR) methods that were also tested when the dataset was sufficiently large. **Figure 8** shows the $R^2$ value of each methodology as the size of the dataset increases. The ROC curve can also be seen along with a graphic of the experimental setup in **Figure 8**. From the wide range of studies published, there are many different applications of fNIRS currently being researched, and the use of DL techniques has become of interest in recent years.

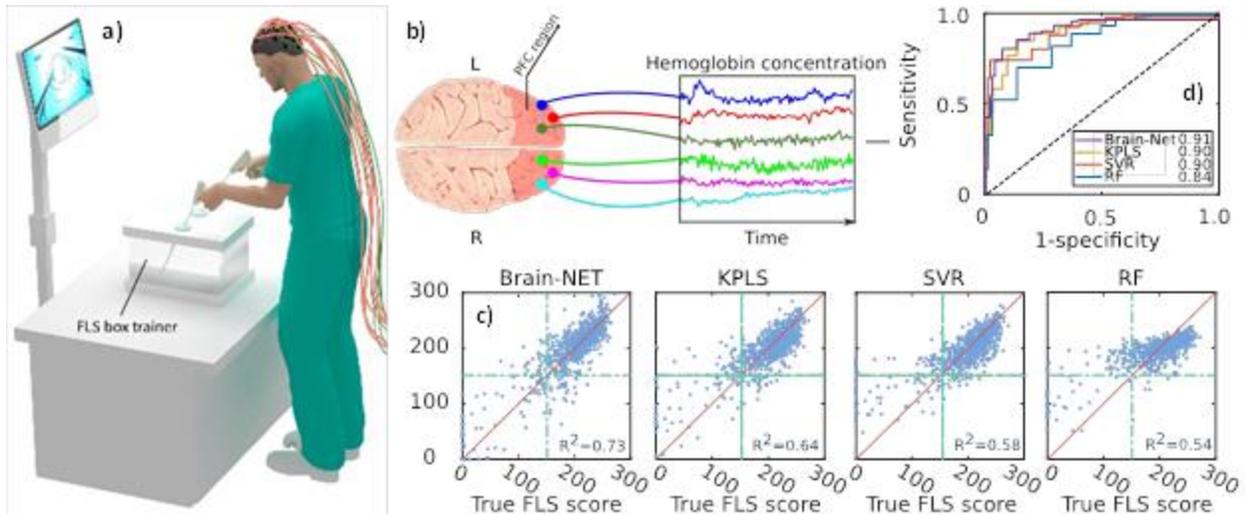

**Figure 8:** (a) Schematic depicting the FLS box simulator where trainees perform the bimanual dexterity task. A continuous-wave spectrometer is used to measure functional brain activation via raw fNIRS signals in real time. (b) Examples of acquired time-series hemoglobin concentration data from six PFC locations while the subject is performing the PC surgical task. (c) The true vs. predicted FLS score plots. Each blue dot represents one sample. The red line is the y = x line. The dot-dashed green lines represent the certification pass/fail threshold FLS score value. (d) ROC curves for each model with corresponding AUC values in the legend. Adapted from Nemani et al. [100] and Gao et al. [32]

## 3 Discussion and future outlook

Over the last two decades, machine learning has become increasingly popular for processing neuroimaging data thanks to its benefit over traditional analysis methods [101]. Of importance, ML methods enable fully processing spatio-temporal data sets and allow for inferencing at the single subject/trial level. Among all ML approaches, DL is becoming increasingly utilized over the last half decade with great promise [102,103]. Following similar trends, DL models have found increased utility in fNIRS applications, ranging from simplifying the data processing pipeline to performing classification or prediction tasks. DL models are expected to outperform ML methods thanks to their potential to directly extract features from raw data (no need to perform prior feature extraction) and learn complex features in a hierarchical manner. This seems to be further supported by the findings of this review in which, out of the 32 papers reporting on a comparative study of deep learning techniques to traditional machine learning techniques, 26 have been shown outperforming the latter in terms of classification accuracy. Such trends have also been reported over a large range of biomedical applications, but we cannot exclude a publication bias due to the specific nature of the review topic. Still, the application of DL to fNIRS is in its very early stages and faces many challenges.

First, the implementation of DL models is an expert field. The selection of the main architecture as well as the number of layers, the activation function, … are still dependent on the user expertise but greatly influence the performance and applicability of any DL model. If this design flexibility enables powerful implementations, it leads to a wide range of architecture and hyperparameters employed in the set of work reviewed herein. Hence, there is still not a consensus on which architecture and hyperparameters are optimal for a specific problem and building on current work requires some level of technical expertise to assess which implementation would be optimal. This may be circumvented in the future with the advent of network automated designed via neural architecture search methods [104], but these have not been yet applied to the field of fNIRS. Moreover, following the principles of the no-free-lunch theorem, it is expected that prior knowledge on the problem at hand should guide in the selection of the ML/DL algorithm. Hence, beyond technical expertise in DL, ones need also to have a good grasp of the neurophysiology as well as instrumentation characteristics used in the application to design optimal models.

This leads to another significant challenge which is associated with the data driven nature of DL model training and validation. The lack of sufficient training data is a common challenge in the application of DL methods in neuroimaging. This is even more challenging for fNIRS applications that are less ubiquitous than MRI or EEG, which benefit from publicly available repository. As we are still far from being able to model the complexity of brain functions and dynamics, the DL models can be trained only on experimental data conversely to many other fields in which efficient *in silico* data generators are available [105]. This is highlighted in Table 3 which shows that almost all reviewed work depended on proprietary data and with relatively small number of subjects. Moreover, in numerous scenarios, the data quality can be poor such that a sub-set of the spatio-temporal data is missing or inadequate (for instance compromised by motion artefacts, shallow physiological variations,…). As previously mentioned, data augmentation approaches have been implemented successfully to alleviate this challenge. Another approach is to leverage new developments in transfer learning that optimally refine well-trained networks on large data sets to smaller one. But still, these methods are expected to work well within homogeneous settings. As the field benefit from an increased number of fNIRS systems with varying optode characteristics and associated electronics, the raw characteristics of the acquired signals can greatly vary (SNR, CNR, sampling rate,…) and hence, limit generalizability. Another issue for generalizability is that overfitting may be especially prevalent in fNIRS studies where fNIRS data tends to be highly correlated [55]. Moreover, in many instances, the data set is imbalanced for available classes. Hence, it is crucial for eliciting confidence in the results to report on Cross Validation (CV) results. Herein, most reviewed work used k-fold CV, typically 5- or 10-fold CV. Still, in many applications, the data set is comprised of multi trials per subject. Hence, it is important to assess the potential bias associated with each subject. This can be performed using Leave-One-User-Out (LOUO) CV and Leave-One-Subject-Out (LOSO) CV respectively. Still, such well-established methods were used only in 8 of the 63 papers considered in this review.

Last, despite demonstrating high performances, the DL implementations reported herein are suffering from the black-box issue. In other words, the extracted high-level features from the data inputs that lead to high task performances during training and validation are not accessible and hence, cannot be interpreted. However, in the last few years, various eXplainable AI (XAI) tools like Class Activation maps [106], Grad-CAMS [107], saliency maps [108] have been proposed to impart understandability and comprehensibility [109]. Such tools can for instance provide visual map(s) that highlight the main data features leveraged for the model decision. Such a map can then correlate the extracted features with known neurophysiology, specific application characteristics (for instance hand switching during surgery) and/or correlate with other biomarkers such as videos, gaze measurements, and motion tracking devices. For this reason, XAI tools are well-poised to lead to the discovery of new spatio-temporal features that will advance our neuroscience knowledge at large. This is exemplified by the recent report of differences in activation maps between expert and novice surgeons while executing a certification task, with activation maps obtained via a dot-attention method in a DL classifier model [110].

## 4. Conclusion

We reviewed the most recent published work relevant to the application of DL techniques to fNIRS. This literature review indicated that DL models were mainly sued for classification tasks based on fNIRS data and that in most of the cases, DL model prediction accuracy outperformed traditional techniques, including established ML methods. Another subset of work reported on developing DL models to reduce the amount of preprocessing typically done with fNIRS data or increase the amount of data via data augmentation. In all cases, DL models provided very fast inference computational times. These characteristics have a transformative power for the field of fNIRS at large as they pave the way to fast and accurate data processing and/or classification tasks. Of note, DL models, when validated and established, offer the unique potential for real-time processing on the bedside at minimal computational cost. While the deployment of DL models that are widely accepted by the community face numerous challenges, the findings reviewed here provide evidence that DL will play an increased role in fNIRS data processing and use for a wide range of bedside applications. Moreover, as an ever-increased number of studies are made available to the community, it is expected that the next generation of DL models will have the possibility to be tested and validated in various scenarios.

**Acknowledgment**

We thank the funding provided by NIH/National Institute of Biomedical Imaging and Bioengineering grants 2R01EB005807, 5R01EB010037, 1R01EB009362, 1R01EB014305, and R01EB019443, the Medical Technology Enterprise Consortium grant 20-05-IMPROVE-004, and the US Army Combat Capabilities Development Command grant W912CG2120001.

**Conflict of interest**

The authors declare that the research was conducted in the absence of any commercial or financial relationships that could be construed as a potential conflict of interest.

[67]     T. K. K. Ho, J. Gwak, C. M. Park and J. -I. Song, "Discrimination of Mental Workload Levels From Multi-Channel fNIRS Using Deep Leaning-Based Approaches," in *IEEE Access*, vol. 7, pp. 24392-24403, 2019, doi: 10.1109/ACCESS.2019.2900127.

[68]     Asgher, U., Khalil, K., Khan, M. J., Ahmad, R., Butt, S. I., Ayaz, Y., Naseer, N., & Nazir, S. (2020). Enhanced Accuracy for Multiclass Mental Workload Detection Using Long Short-Term Memory for Brain-Computer Interface. *Frontiers in neuroscience*, *14*, 584. https://doi.org/10.3389/fnins.2020.00584

[69]     Noman Naseer, Nauman Khalid Qureshi, Farzan Majeed Noori, Keum-Shik Hong, "Analysis of Different Classification Techniques for Two-Class Functional Near-Infrared Spectroscopy-Based Brain-Computer Interface", *Computational Intelligence and Neuroscience*, vol. 2016, Article ID 5480760, 11 pages, 2016. https://doi.org/10.1155/2016/5480760

[70]     Hakimi, N., Jodeiri, A., Mirbagheri, M., & Setarehdan, S. K. (2020). Proposing a convolutional neural network for stress assessment by means of derived heart rate from functional near infrared spectroscopy. *Computers in Biology and Medicine*, *121*, 103810. https://doi.org/10.1016/j.compbiomed.2020.103810

[71]     Erdoĝan, S. B., Özsarfati, E., Dilek, B., Kadak, K. S., Hanoĝlu, L., & Akın, A. (2019). Classification of motor imagery and execution signals with population-level feature sets: implications for probe design in fNIRS based BCI. *Journal of neural engineering*, *16*(2), 026029. https://doi.org/10.1088/1741-2552/aafdca

[72]     Hamid, H., Naseer, N., Nazeer, H., Khan, M. J., Khan, R. A., & Shahbaz Khan, U. (2022). Analyzing Classification Performance of fNIRS-BCI for Gait Rehabilitation Using Deep Neural Networks. *Sensors (Basel, Switzerland)*, *22*(5), 1932. https://doi.org/10.3390/s22051932

[73]     Khan, H., Noori, F. M., Yazidi, A., Uddin, M. Z., Khan, M., & Mirtaheri, P. (2021). Classification of Individual Finger Movements from Right Hand Using fNIRS Signals. *Sensors (Basel, Switzerland)*, *21*(23), 7943. https://doi.org/10.3390/s21237943

[74]     Ortega, P., & Faisal, A. A. (2021). Deep learning multimodal fNIRS and EEG signals for bimanual grip force decoding. *Journal of neural engineering*, *18*(4), 10.1088/1741-2552/ac1ab3. https://doi.org/10.1088/1741-2552/ac1ab3

[75]     Q. Zhao, C. Li, J. Xu, and H. Jin, "FNIRS based brain-computer interface to determine whether motion task to achieve the ultimate goal," 2019, doi: 10.1109/ICARM.2019.8833883.

[76]     H. Ghonchi, M. Fateh, V. Abolghasemi, S. Ferdowsi, and M. Rezvani, "Spatio-temporal deep learning for EEG-fNIRS brain computer interface," in *Proceedings of the Annual International Conference of the IEEE Engineering in Medicine and Biology Society, EMBS*, 2020, vol. 2020-July, doi: 10.1109/EMBC44109.2020.9176183.
38